# Distributed Frequency Emergency Control with Coordinated Edge Intelligence

Yingmeng Xiang[1], Zhehan Yi[1], Xiao Lu[2], Zhe Yu[1], Di Shi[1], Chunlei Xu[2], Xueming Li[3] and Zhiwei Wang[1]

[1] GEIRI North America, San Jose, CA 95134 USA
[2] Dispatch Center, State Grid Jiangsu Electric Power Company Ltd, Nanjing, Jiangsu, China
[3] NARI Group Corporation, Nanjing, China

Abstract—Developing effective strategies to rapidly support grid frequency while minimizing loss in case of severe contingencies is an important requirement in power systems. While distributed responsive load demands are commonly adopted for frequency regulation, it is difficult to achieve both rapid response and global accuracy in a practical and cost-effective manner. In this paper, the cyber-physical design of an Internet-of-Things (IoT) enabled system, called Grid Sense, is presented. Grid Sense utilizes a large number of distributed appliances for frequency emergency support. It features a local power loss $\Delta P$ estimation approach for frequency emergency control based on coordinated edge intelligence. The specifically designed smart outlets of Grid Sense detect the frequency disturbance event locally using the parameters sent from the control center to estimate active power loss in the system, and to make rapid and accurate switching decisions soon after a severe contingency. Based on a modified IEEE 24-bus system, numerical simulations and hardware experiments are conducted to demonstrate the frequency support performance of Grid Sense in the aspects of accuracy and speed. It is shown that Grid Sense equipped with its local $\Delta P$-estimation frequency control approach can accurately and rapidly prevent the drop of frequency after a major power loss.

## 1. Introduction

Power system frequency deviation can be caused by a mismatch between load demand and power generation. This may be due to a number of factors, including loss of generators, tripping of transmission lines, fluctuation of load demand, variation in renewable generation, to name a few. In recent years, with increased penetration of power-electronic-interfaced generators, intermittency of renewable generation, as well as potential natural disasters and cyber-physical threats [1-3] the demand to maintain frequency stability has become all the more challenging. In some countries, e.g., China, there are many load centers that are powered by multiple ultrahigh voltage direct current (UHVDC) or ultrahigh voltage alternative current (UHVAC) lines [4]. These lines have very high transmission capacity; if a contingency occurs in one, it can cause severe power shortages and pose grave threats to the power system's stability.

In real terms, there have been several severe blackouts due to frequency disturbances. On Aug. 8, 2019, for example, the gas-fired power plant, Little Barford, and the offshore wind farm, Hornsea, both in the UK were tripped, resulting in severe frequency drops and a power outage that affected 1.1 million houses and businesses for about an hour [5]. In March 2018, the failure of a UHVDC line in Brazil caused a massive power outage, impacting tens of millions of people [6].

Power system frequency regulation is conventionally achieved by means of a hierarchical system that includes primary, secondary, and tertiary frequency controls. However, these controls are usually unable to prevent blackouts in case of a sudden major power-loss contingency. In recent years, demand response and direct load control have been proposed and studied as an alternative to frequency regulation, as reviewed in [7]. However, most of these approaches are slow (hourly or minutes) and designed only for long-term demand response and peak load shaving. Also, the wide deployment of the direct load control in these approaches is substantially hindered by the degree of investment needed for installation and modification of both the communication framework and the control hardware [8-10].

Generally, for widely distributed responsive loads to participate in frequency regulation, there are two methods: centralized control and decentralized control [7]. In a centralized control scheme, some researchers [11-12] have suggested adaptive approaches while others utilize a $\Delta P$-estimation [13-15]. In adaptive approaches [11-12], the central controller measures the power system frequency, and calculates the required load curtailment using methods such as PI control and droop control. For the $\Delta P$-estimation [13-15], the central controller first estimates the magnitude of the power loss $\Delta P$ after the disturbance, and then calculates the load curtailment based on the $\Delta P$-estimation. While these centralized control schemes may differ in the calculation of load curtailments, a common feature they share is that the central controller sends the control signals to the distributed response loads. The centralized control strategy has the advantages of global accuracy, but it requires a large number of costly communication channels, which may result in a slow response of distributed devices, thereby making it unsuitable for frequency emergency control. For the use case of UHVDC fault, the power system frequency usually drops significantly within a few seconds, requiring a rapid response of a large array of devices, which are sparsely distributed geographically in a city or in a province. The centralized control, therefore, has many challenges in dealing with such faults.

There also exist a range of decentralized control approaches [16-18]. Many of these share common factors, such as states of distributed loads, frequency thresholds, and time delay [16-18], which allow distributed control devices to make autonomous switching decisions. The advantage of decentralized control is that it does not rely on communication and can respond quickly. Nevertheless, without centralized coordination, it can usually lead to insufficient or excessive load curtailment. This is because without system-level parameters such as the power system inertia constant, the local frequency measurement cannot reveal the amount of load shedding needed to regulate the disturbance.

In power systems, there can be some great faults, e.g., UHVDC line tripping, or sudden tripping of great generators. When such a great fault happens, the power system frequency can drop to a dangerous level rapidly within a few seconds, like in the UK blackout [5]. In such cases, the centralized methods might be slow as it depends on communication. The existing decentralized methods are also slow as they passively wait until the frequency drops below a threshold. How can we develop a method which can respond to such great faults actively, accurately, and rapidly?

In this paper, a novel load control system called Grid Sense is developed to achieve real-time monitoring and direct control of end-user load appliances [19]. The Grid Sense system is a distributed system using an extensive array of controllable loads for frequency control, consisting mainly of smart outlets, and a control center in the cloud. Furthermore, based on this system, a local ΔP-estimation approach for frequency emergency control is proposed and implemented. Compared with the aforementioned centralized and decentralized frequency control methods, the proposed local ΔP-estimation approach in this paper has the following advantages: (1) Different from a centralized approach, which estimates power loss in the central controller, the proposed local ΔP-estimation approach estimates power loss locally in each of the distributed control devices and makes switching decisions. In this way, it combines some merits of the centralized control and decentralized control, such as global accuracy and fast response. (2) The distributed loads are ranked from least important to most important, and an index named accumulated power is proposed, which indicates load shedding priorities. The switching of a distributed control device is based on comparing power loss estimation and accumulated power, and this makes the switching of the distributed smart outlet coordinated with each other. Further, the overall load shedding in the power system can be accurate even if there are some errors in the power loss estimation of some individual outlets.

The contributions of this paper are summarized as follows.

(1) The cyber-physical architecture and implementation of an industrial system that is designed for frequency support under UHVDC faults, namely Grid Sense, is elaborated.

(2) A novel distributed frequency control approach, namely local ΔP-estimation, is proposed. This approach is based on coordinated intelligence of the distributed edge devices to enable a large number of widely distributed appliances to provide fast and robust frequency support in case of a severe contingency.

(3) Numerical simulations, hardware experiments, and comparative studies are conducted to verify the accuracy and robustness of the frequency control strategy of the Grid Sense system.

The rest of this paper is organized as follows. Section II briefly recaps the power system frequency dynamics model. The development of Grid Sense and the overview of the adopted frequency control approaches are presented in Section III. The control center strategy of the control center is explained in detail in Section IV, while the control strategy of the smart outlets is elaborated in Section V. Case studies are performed in Section VI. Conclusions are provided in Section VII.

## 2. Modeling of Power Grid Frequency Dynamics

In power system operations, it is critical to ensure the balance between load and generation at any time. When a major contingency occurs, such as sudden loss of a large generator or an interconnection tie line, this may cause power mismatch to occur, which results in frequency fluctuations. To model the power system frequency dynamics in response to power imbalance, a simplified system frequency response model developed in [20] is widely adopted, and explained as follows.

$$\Delta P_G(t) - \Delta P_D(t) - D\Delta f(t) = 2H\frac{d\Delta f(t)}{dt} \quad (1)$$

where $\Delta P_G(t)$ is the total increase of power generation in per unit (p.u.) at time $t$; $\Delta P_D(t)$ is the total increase of load demand (p.u.) at time $t$; $D$ is the amount of load damping, which indicates the sensitivity of the load in response to frequency change; $\Delta f(t)$ is the frequency change (p.u.); $H$ is the equivalent inertia constant of the system. For a system with multiple generators, the value of $H$ can be obtained as follows:

$$H = \frac{1}{S_B}\sum_{i=1}^{N} H_i S_{B,i} \tag{2}$$

where $S_B$ is the base power of the system; $S_{B,i}$ is the power rating of the $i^{th}$ generator; $H_i$ is the inertia constant of the $i^{th}$ generator; $N$ is the total number of the generators in operation, and the tripped generators should be excluded.

The value of $\Delta P_G(t)$ can change due to the control of the generator governors, which can be described as in Fig. 1.

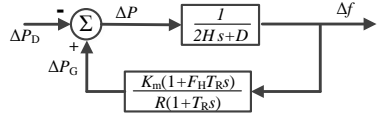

Fig. 1 Frequency response model of the power system

In Fig. 1, $\Delta P_{SP}$ is the change of the generator power setpoint; $F_H$ is the fraction of the power generator by the reheat turbine; $T_R$ is the average reheat time constant; $K_m$ is the power gain factor; $R$ is a constant of the governor speed-droop control. Typically, changing the generator power setpoint takes time, and thus $\Delta P_{SP}$ is ignored in most emergency cases.

When a power mismatch between the generation and load demand occurs, the system frequency will experience a dynamic change until a new equilibrium is reached. The frequency response to the power imbalance can be calculated using the Laplace transform as follows:

$$\Delta f(s) = \frac{R\omega_n^2}{DR+K_m} \cdot \frac{1+T_R s}{s^2+\zeta\omega_n s+\omega_n^2} \cdot \frac{\Delta P}{s} \tag{3}$$

where

$$\omega_n = \sqrt{\frac{DR+K_m}{2HR \cdot T_R}} \tag{4}$$

$$\zeta = \frac{2HR+(DR+K_m F_H)T_R}{2(DR+K_m)}\omega_n \tag{5}$$

The frequency response in the time domain can be computed by the inverse Laplace transform, as [20]

$$\Delta f(t) = -\frac{R\Delta P}{DR+K_m}[1 + \alpha e^{-\zeta\omega_n t}\sin(\omega_r t + \emptyset)] \tag{6}$$

where

$$\alpha = \sqrt{\frac{1-2T_R\zeta\omega_n+T_R^2\omega_n^2}{1-\zeta^2}} \tag{7}$$

$$\omega_r = \omega_n\sqrt{1-\zeta^2} \tag{8}$$

$$\emptyset = \emptyset_1 - \emptyset_2 = tan^{-1}\left(\frac{T_R\omega_r}{1-\zeta\omega_n T_R}\right) - tan^{-1}\left(\frac{\sqrt{1-\zeta^2}}{-\zeta}\right) \tag{9}$$

When there are multiple sudden power changes $\Delta P(t_j)$ at different times $t_j$, the frequency response is as follows:

$$\Delta f(t) = -\sum_j \frac{R\Delta P(t_j)}{DR+K_m}\{1 + \alpha e^{-\zeta\omega_n(t-t_j)}\sin[\omega_r(t-t_j)+\emptyset]\} \tag{10}$$

## 3. Development of Grid Sense and Overview of Its Frequency Control Approaches

In order to curb the rapid drop in frequency after a UHVDC fault in China, a direct load control system, such as the Grid Sense system is developed. Its major architecture and frequency control approaches are as follows:

*3.1 Introduction to Grid Sense*

The overall cyber-physical architecture of Grid Sense is displayed as in Fig. 2. The Grid Sense system mainly consists of the following major components:
- ➢ Control center: to receive, analyze, and display the measurements from the distributed smart outlets, and to broadcast parameters and commands to the smart outlets to control their operations.
- ➢ Communication network: to interface with the distributed smart outlets and control.
- ➢ Smart outlets: to monitor and control connected devices in real time.
- ➢ Devices: a variety of home and commercial devices, including air conditioners, refrigerators, EV chargers, lamps, and heaters can be easily plugged into smart outlets by the customers. These appliances serve as responsive load demands.

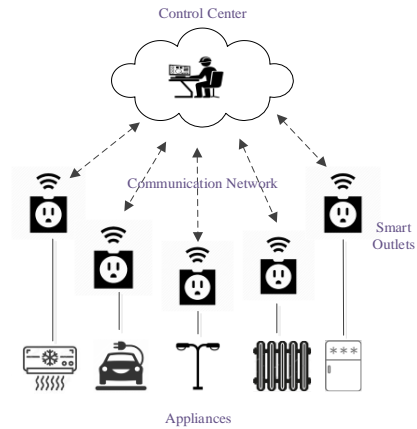

Fig. 2 Cyber-physical architecture of the Grid Sense system

Specifically, the smart outlets with edge intelligence play a key role in the Grid Sense system. They can: 1) measure voltage, current, active power, and reactive power; 2) track dynamic frequency in real time; 3) send measurements to the cloud-based control center; 4) control the switch based on local measurements or commands from the control center; and 5) receive settings from the cloud, including time delay and thresholds. An extended Kalman filter (EKF) method is developed and adopted for fast frequency tracking [19], and the EKF method delivers a frequency measurement result every 16 milliseconds continuously.

The function blocks of the smart outlet are depicted in Fig. 3.

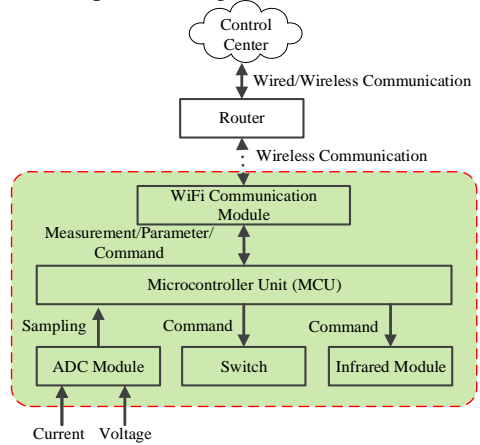

Fig. 3. Function blocks of the smart outlet

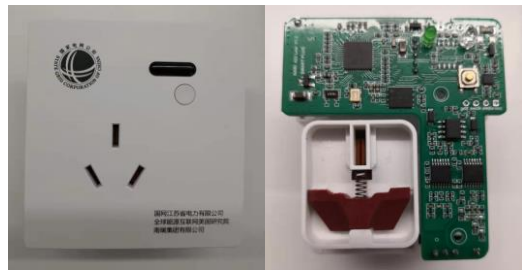

Fig. 4. Development of smart outlets

The hardware of the smart outlets has been developed and tested, as shown in Fig. 4. WiFi is used for communication between a smart outlet and router. When compared with wired communication, such as fiber optics, WiFi communication provides a low-cost and flexible solution. Furthermore, the communication between the control center and different geographically distributed smart outlets can be achieved using the MQTT protocol, which is a lightweight messaging protocol commonly used in Internet of Technology (IoT) systems.

A cloud center is set up to receive measurements from the smart outlets, and to detect smart outlet locations, as well as store, analyze, and contour the measurements on a map. The cloud is also used to broadcast settings or control commands to smart outlets. To facilitate power system operators, a user-friendly interface is developed, as shown in Fig. 5.

The Grid Sense system, as explained here, is currently being implemented in State Grid Corporation of China (SGCC) Jiangsu, one of the largest provincial power companies in China.

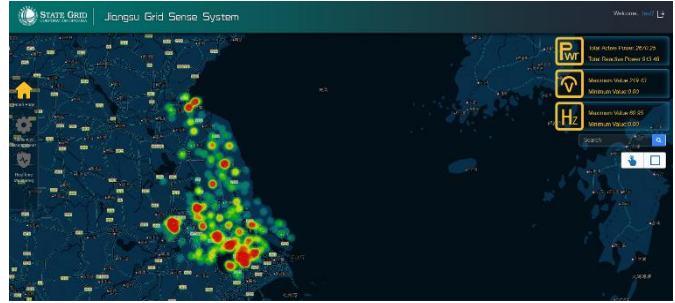

Fig. 5 Control center of the Grid Sense system

*3.2 Overview of Frequency Control Approaches*

In the Grid Sense system, a holistic frequency control strategy, including three schemes are implemented. The functions and the coordination of each are elaborated as follows:

(1) *Local ΔP-estimation approach:* Before any frequency disturbance event can occur, the cloud control center monitors the status of the smart outlets, calculates the parameters of the power system (e.g., the inertia time constant, load damping factor)for the smart outlets to perform ΔP-estimation locally, and sends the parameters to the smart outlets. As the power system state may change with time, the parameters will be updated by the control center and sent to the smart outlet periodically, e.g., every 15 minutes. The smart outlets then update and store the parameters in the local memory. Based on these parameters, the smart outlets then is able to detect frequency disturbance events, estimate occurrence of power loss, and make decisions to cut loads if certain criteria are met. Based on this local ΔP-estimation approach, the smart outlets usually take load shedding actions within 1 second when needed, in order to effectively curb frequency drop in case of a major frequency disturbance event.

(2) *Adaptive switching-off approach:* Before a frequency disturbance event, the control center sets a switching-off frequency for each smart outlet and sends it to these outlets. If a smart outlet fails to respond based on the local ΔP-estimation approach, the smart outlet will immediately switch off the load connected to it when the measured frequency is lower than the switching-off frequency. The decision-making of the smart outlets using the adaptive switching-off approach is dependent on the frequency measurement. Due to the inertia in the power system, the frequency usually drops to the switching-off frequency after a few seconds. Thus, the adaptive switching-off approach usually takes actions later than the local ΔP-estimation approach.

(3) *Command-based direct load shedding approach:* After the frequency disturbance event, the control center will estimate the amount of load shedding needed and monitor the response of the smart outlets. The control center can send the commands to directly switch on/off the smart outlets if some of the outlets fail to respond or respond in error.

The local ΔP-estimation approach serves as the primary frequency control approach. It has the advantage of both fast-response and global accuracy while the communication burden is greatly reduced. This is explained in detail in Sections IV, V, and VI. The other two approaches, i.e., adaptive switching-off approach and command-based direct load shedding approach, serve as backup approaches.

In the rest of the paper, the local ΔP-estimation approach is explained in detail from the perspectives of the control center and the smart outlets. Since the adaptive switching-off approach and the command-based direct load shedding approach are widely studied, they will only be briefly mentioned.

## 4. Frequency Control Strategy in the Control Center

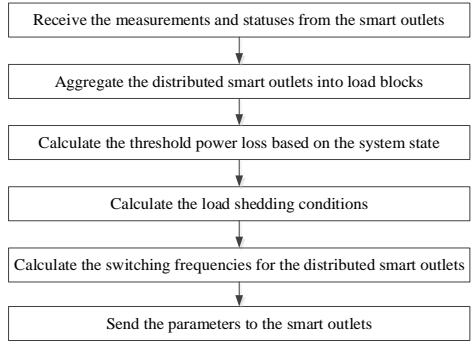

Fig. 6  Control strategy in the cloud

The control center's main functions, as shown in Fig. 6, are set up to calculate the parameters for coordinating the distributed smart outlets switching system. These parameters are then sent to the smart outlets before a frequency disturbance event. Since the parameters are calculated and sent before the event, the outlets can tolerate communication delays, while at the same time reduce the communication burden.

*4.1 Aggregation of Smart Outlets*

The smart outlets measure in real time such functions as voltage, current, active power, reactive power, frequency, rate of change of frequency (ROCOF), and the switch status of plugged-in appliances. The smart outlets also send these measurements to the cloud at 1-minute intervals to update the control center on their status. The control center aggregates the widely distributed smart outlets into hierarchical load blocks, taking into consideration the types of appliances and their locations, as shown in Fig. 7.

Minimizing the impact of load shedding to the customers is a critical concern of Grid Sense. The load shedding should be conducted carefully, and unnecessary load shedding should be avoided. In general, for customers, some loads are more important than others. For example, air conditioners are typically less important than computers. Load shedding near the receiving end of a UHVDC line can in relative terms more effectively prevent the power flow transfer and alleviate the transient stability issues. Based on the types and locations of the load blocks, they can be sorted from the least to the most important. Note that the operators in the control center can have different criteria to sort and rank the loads, and it is not the focus of the proposed work.

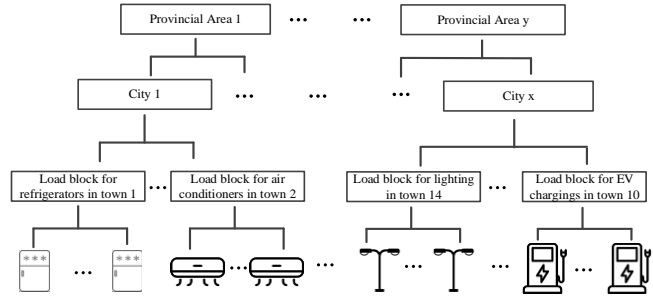

Fig. 7 Aggregation of the loads

Assume the power of the $i^{th}$ block is $P_i^B$, thus the accumulated power for the ith block is defined as
$$P_i^A = \sum_{j=1}^{i} P_j^B \tag{11}$$

The accumulated power of each block can be calculated using (11). The accumulated power of a load block indicates the priority of that load block in the event of load shedding and plays a very important role in ensuring the robustness of the proposed local $\Delta P$-estimation approach, which will be explained in detail later.

The accumulated power of each block will be sent to the smart outlets in that block. After a frequency disturbance event, the smart outlet will estimate the power loss. The smart outlet will then switch off only if the amount of required load shedding is greater than the value of accumulated power stored in the smart outlet.

The blocks serve as the basis for the control and management of the smart outlets. It is noteworthy that the blocks are not fixed and they can change, i.e., merge or divide dynamically as needed.

*4.2 Determination of Threshold Power Loss*

During the power system's long-term operation, there will be fluctuations in frequency. Generally, mild fluctuations that do not cause significant damage to the power system can be taken care of by primary or secondary frequency regulation. The Grid Sense system should not curtail load in such mild cases.

One principle of the Grid Sense system is to ensure that the system frequency does not drop to the point of under frequency load shedding (UFLS) while minimizing the amount of load shed by the smart outlets. In this way, potential system collapse can be prevented, thus avoiding large-scale load shedding by UFLS and allowing time for other frequency control methods to take actions and restore the power system frequency.

The starting frequency of UFLS is different for different systems. In China, for a 50 Hz system, a typical value of UFLS starting frequency is 49 Hz. We choose a threshold frequency slightly higher than the starting frequency of the UFLS, denoted as $f_s$, as the objective for the frequency control of Grid Sense. A typical value of $f_s$ can be 49.5 Hz. In other words, the objective of the local $\Delta P_L$-estimation approach is to enable the smart outlets to respond rapidly in case of a major frequency disturbance and ensure that the frequency does not drop below $f_s$.

For a given power system state, we define the amount of threshold power loss as $\Delta P_s$ which makes the system frequency drop to $f_s$ at a minimum. The calculation of the $\Delta P_s$ is explained as follows.

The response to a sudden loss of a generator or increase in load in typical power system frequency dynamics is shown in Fig. 8. The frequency decreases until the minimum frequency $f_{nadir}$ is reached. Then the frequency goes up until a new equilibrium is achieved.

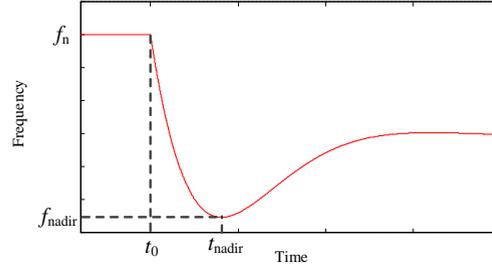

Fig. 8 Typical system dynamics following a power loss

According to (6), given the initial power imbalance $\Delta P$ at time $t_0=0$, the ROCOF $g(t)$ can be calculated as

$$g(t) = \frac{d\Delta f(t)}{dt} = -\frac{\alpha \omega_n R \Delta P}{DR+K_m} e^{-\zeta \omega_n t} \sin(\omega_r t + \emptyset_1)] \tag{12}$$

The frequency $f_{nadir}$ is reached when the $ROFC$ is zero. Thus, the time $t_{nadir}$ needed to reach $f_{nadir}$ can be calculated as follows

$$t_{nadir} = \frac{\pi - \emptyset_1}{\omega_r} = \frac{1}{\omega_r} tan^{-1}(\frac{T_R \omega_r}{\zeta \omega_n T_R - 1}) \tag{13}$$

Therefore, the $f_{nadir}$ can be obtained using the following equation [13].

$$f_{nadir} = f_n - \frac{R\Delta P}{DR+K_m}[1 + \alpha e^{-\zeta \omega_n t_{nadir}} \sin(\omega_r t_{nadir} + \emptyset)] \tag{14}$$

where $f_n$ is the pre-contingency frequency of the power system in p.u. and it is typically 1.

Hence, based on (12) the threshold power loss $\Delta P_s$ which makes the frequency drop to $f_s$ as the nadir frequency can be calculated as

$$\Delta P_s = \frac{(f_n - f_s) \times (DR+K_m)}{R \times [1+\alpha e^{-\zeta \omega_n t_{nadir}} \sin(\omega_r t_{nadir}+\emptyset)]} \tag{15}$$

*4.3 Calculation of Load Shedding Conditions*

When a major contingency occurs and causes the power system frequency to drop significantly, the smart outlets need to respond quickly to cut off the load connected to them in order to support the frequency. During the frequency drop process following a major contingency, the frequency goes through complex dynamics. Specifically, there can be various noises, harmonics, distortion, and spikes at the smart outlets, which are located at the user-end level of the power system,. A robust operation strategy, therefore, is required to enable the smart outlets to perform accurately in noisy environments; absent a robust strategy, there can be mis-operation, which can lead to undesired tripping of the loads, causing inconvenience to customers or even worsening the situation.

In this part, a robust and adaptive load shedding condition is proposed for the smart outlets to adaptively determine whether load shedding is needed and to avoid mis-operations.

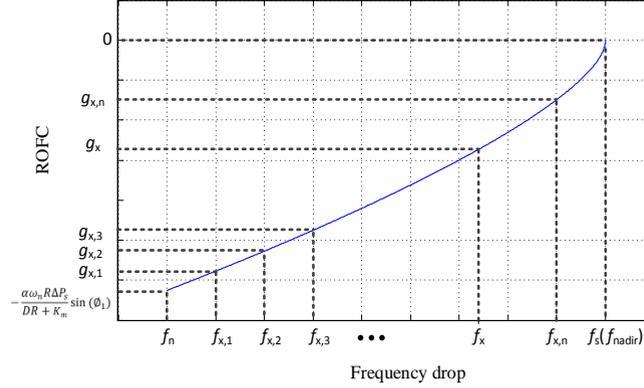

Fig. 9 The relationship between the ROCOF and frequency drop

Based on $\Delta P_s$ calculated using (15), the frequency of the power system following the power loss is calculated using (6). Also, the ROCOF can be obtained using (12). A typical relationship between the ROCOF and the frequency is represented as shown in Fig. 9. Immediately after the loss of power $\Delta P$ at time $t_0$, the ROCOF $g(t_0)$ is

$$g(t_0) = -\frac{\alpha \omega_n R \times \sin(\emptyset_1)}{DR + K_m} \Delta P \qquad (16)$$

It is noted that the ROCOF value is negative; thus the ROCOF will grow larger (i.e, the speed of the frequency drop is decreasing) until it reaches 0 when the frequency is decreased to $f_s$. For any frequency value $f_x$ between $f_n$ and $f_s$, a corresponding ROCOF value $g_x$ can be found. If the real-time frequency measurement is $f_x$ and at the same time the ROCOF measurement is lower than $g_x$, the switching-off condition is satisfied for one time. For robustness, the smart outlet will decide that load shedding in the power system is needed only when the switching-off condition is satisfied for multiple times.

Next, the calculation of load shedding conditions is elaborated considering the practical factors. In the curve shown in Fig. 9, there can be an infinite number of $(f_x, g_x)$ pairs. Since it is not realistic to store every pair, as the storage within the smart outlet is limited and the control strategy should not be too complicated, the frequency range between $f_n$ and $f_s$ is divided into pieces.

*4.4 Calculation of Switching-off Frequency*

For practical considerations, the adaptive switching-off approach is implemented as a backup control approach in case a few of the millions of smart outlets fail to respond based on the local $\Delta P$-estimation approach. For each load block, a switching-off frequency $f_i^B$ is calculated as follows.

$$f_i^B = f_s - \frac{P_i^B}{D(P_L - \sum_{i=1}^{i-1} P_i^B)} f_n \qquad (17)$$

where $P_L$ is the total amount of load demand in the power system.

The switching-off frequency of a load block is sent to all the smart outlets in that load block. A smart outlet will switch off immediately once the measured frequency is lower than the corresponding switch-off frequency.

**5. Control Strategy in Smart Outlets**

According to the holistic frequency control strategy introduced above, the flowchart of the smart outlets abiding by these three approaches is shown in Fig. 10.

In the flowchart, steps 5-9 correspond to the local ΔP-estimation approach. Step 10 corresponds to the adaptive switching-off approach. Step 11 corresponds to the command-based direct load shedding approach.

Several key steps, including abnormal frequency disturbance event detection, load shedding condition check, power loss estimation using least square estimation (LSE), and power loss estimation using EKF are explained in detail as follows.

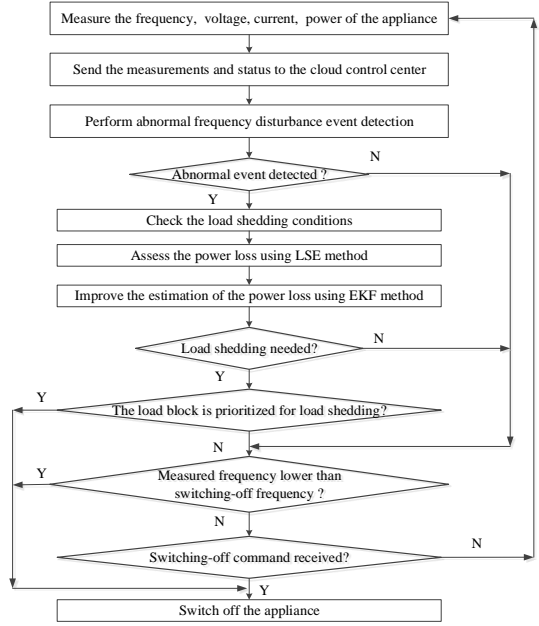
Fig. 10 Flowchart of a smart outlet

## 5.1 Abnormal Frequency Distribution Event Detection

When there is a change in frequency measurement, it is critical to check whether the change is a credible contingency or just normal fluctuations. To ascertain this, an abnormal frequency disturbance event detection method is needed.

To measure power system frequency in real time, the smart outlet utilizes an EKF-based method, which provides a frequency measurement result every 16 milliseconds. While it may seem reasonable to use a sudden frequency drop to detect an abnormal event, in noisy environments at the end of a distribution system where the smart outlets are located, this should be conducted carefully because if a single point of frequency drop is used, a high probability of false detection can occur. For example, if the noise causes a frequency measurement deviation of 0.01 Hz, the smart outlet may estimate that the frequency drop rate is 0.01Hz/0.016s=0.625 Hz/s, which is a high rate of frequency drop.

For a sudden power loss, like a UHVDC line tripping or a major generator trip, the power system frequency will drop continuously. Thus, the abnormal event detection method in the smart outlets will be as follows: the smart outlet determines that an abnormal frequency disturbance event exists *only* when the frequency drops continuously for five consecutive frequency measurements. In this way, the frequency disturbance event is more robust in case of random noises.

## 5.2 Load Shedding Condition Check

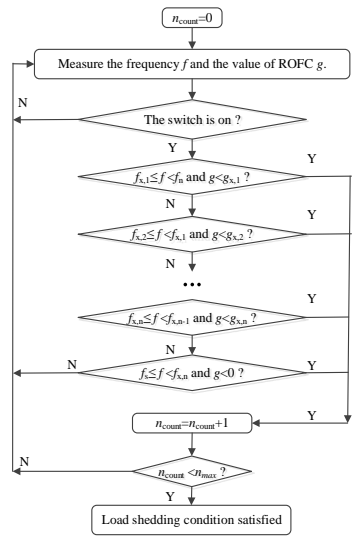

Immediately after an abnormal frequency disturbance event is detected, the smart outlet checks to see whether any of the loading shedding conditions within the rows as seen in Fig. 9 is met for each frequency measurement obtained every 16 milliseconds. To increase the robustness of this method, the smart outlet decides that load shedding is needed only when $n_{max}$ frequency measurement satisfy the conditions in Fig. 9, as shown in Fig. 11.

*5.3 Power Loss Estimation Using LSE*

According to (16), the value of power loss $\Delta P$ can be estimated according to the ROCOF immediately after power loss $g(t_0)$, as

$$\Delta P = -\frac{DR+K_m}{\alpha\omega_n R \times \sin(\emptyset_1)} g(t_0) \tag{18}$$

After an abnormal frequency disturbance event is detected, a few frequency measurements are adopted to estimate the $g(t_0)$. Assume $n$ frequency measurements after the disturbance as $FY = \{f_1, f_2, \cdots, f_n\}$, denote $FX = \{1,2,\cdots,n\}/f^s$, where $f^s$ is the reporting rate of the frequency measurements.

$$g(t_0) = \frac{\sum_{i=1}^{n}(FY_i - \overline{FY})(FX_i - \overline{FX})}{\sum_{i=1}^{n}(FX_i - \overline{FX})^2} \tag{19}$$

where $\overline{FY}$ and $\overline{FX}$ are the average values of $FY$ and $FX$, respectively.

It is noted that the number of frequency measurements used to estimate $g(t_0)$ should not be too small or too large. If the number is too small, the result of $g(t_0)$ might be greatly affected by the possible errors in a few measurements since there can usually be significant harmonics/noises in the power system after a sudden significant power loss. It should not be too large as well because (18) is only valid after a sudden power loss. After sufficient tests, we find 10 measurements can be a good choice.

Denote $K^{LSE} = -\frac{DR+K_m}{\alpha\omega_n R \times \sin(\emptyset_1)}$, which is a constant for a given power system state, and can be calculated before the event by the control center and sent to the smart outlet. Based on the $K^{LSE}$ and the estimated value of $g(t_0)$, an initial estimate of the power loss several milliseconds after the power loss can be obtained as $\Delta P^{init} = K^{LSE} \times g(t_0)$.

*5.4 Power Loss Estimation Improvement Using EKF*

An EKF method is adopted to further improve the estimation accuracy, and the loss estimated by the LSE serves as an initial solution for the EKF.

After a sudden power loss $\Delta P(t_x)$ at $t_x$, according to the power system frequency dynamics model in Section II, the frequency will be

$$f(t) = f_n - \frac{R\Delta P(t_x)}{DR+K_m}\{1 + \alpha e^{-\zeta\omega_n(t-t_x)}\sin[\omega_r(t-t_x) + \emptyset]\} \tag{20}$$

In practice, for distributed devices such as smart outlets, which continuously measure the frequency, it is very challenging to know the exact value of $t_x$ in real time. Also, there is the consideration that the power loss based on a wrong value of $t_x$ using (18) will greatly affect the estimation of $\Delta P$.

Thus, there are two values that need to be estimated here, namely, $\Delta P$ and $t_x$. In other words, if we want to estimate the power loss using the frequency measurements after some time of the power loss, we need to simultaneously estimate the amount of power loss, together with the time of the power loss.

Considering the reporting rate $f^s$ of the frequency measurements, equation (20) is discretized as (21).

$$f(k) = f_n - \frac{R\Delta P(t_x)}{DR+K_m}\left\{1 + \alpha e^{-\frac{\zeta\omega_n}{f^s}(t(k)-t_x)}\sin\left[\frac{\omega_r}{f^s}(t(k)-t_x) + \emptyset\right]\right\} \tag{21}$$

where $k$ is the index of the EKF iterations.

Define the state for the EKF as

$$X[k] = \begin{bmatrix} \Delta P \\ t_x \end{bmatrix} \tag{22}$$

The state transition equation is as follow:

$$X[k+1] = \begin{bmatrix} 1 & 0 \\ 0 & 1 \end{bmatrix} X[k] \tag{23}$$

Then, the observation matrix is linearized as follows:

$$H = \begin{bmatrix} G_1\left\{1 + \alpha e^{-\frac{\zeta\omega_n}{f^s}(t(k)-t_x)} \sin(G_3)\right\} \\ G_2\Delta P\left\{\frac{\zeta\omega_n}{f^s}\sin(G_3) - \frac{\omega_r}{f^s}\cos(G_3)\right\} \end{bmatrix} \quad (24)$$

$$G_1 = -\frac{R}{DR+K_m} \quad (25)$$

$$G_2 = G_1 \alpha e^{-\frac{\zeta\omega_n}{f^s}(t(k)-t_x)} \quad (26)$$

$$G_3 = \frac{\omega_r}{f^s}(t(k) - t_x) + \emptyset \quad (27)$$

The observation equation is

$$f(k) = H \times X[k] + f_n + \varepsilon[k] \quad (28)$$

where $\varepsilon[k]$ is the noise.

By using the EKF method for each frequency measurement in real time, a smart outlet can locally estimate the power loss. More details about the EKF method can be found in [21].

Generally, the accuracy of the EKF relies heavily on the selection of the initial estimation. To alleviate this problem, the power loss obtained by the LSE method serves as the initial estimation of the EKF, which helps to improve the accuracy of the EKF method.

If the estimated power loss is $\Delta P$, the minimal load shedding requirement $\Delta P_v$ that needs to shed in order to prevent the frequency from dropping below $f_s$ is

$$\Delta P_v = \begin{cases} \Delta P - \Delta P_s & \text{if } \Delta P > \Delta P_s \\ 0 & \text{if } \Delta P \leq \Delta P_s \end{cases} \quad (29)$$

*5.5 Switching-off of Smart Outlets*

In sum, a smart outlet will switch off if any of the following three criteria are met.

(1) The smart outlet detects a major power system disturbance event, and estimates that the minimal load shedding requirement $\Delta P_v$ in (29) is larger than the value of the accumulated power of the load block in (11). This corresponds to the local $\Delta P$-estimation approach.

(2) The measured frequency is below the switching-off frequency $f_i^B$ shown in (17). This corresponds to the adaptive switching-off approach.

(3) The smart outlet receives a switching-off command from the control center. This corresponds to the command-based direct load shedding approach.

**6. Case Studies**

Case studies are carried out on a modified IEEE 24-bus system to demonstrate the merits of the frequency control approaches of the Grid Sense system. The topology of the modified IEEE 24-bus system is shown in Fig. 12, in which the generators on bus 23 are removed and a UHVDC line is connected to bus 23. The modified IEEE 24-bus system is adopted to represent the characteristics of a few power grids in China where there are UHVDC lines with capacities much greater than the output of a generator.

The major parameters used in the model for power grid frequency dynamics analysis are as follows. For those generators with a rated power of less than 100 MW, the inertia constant $H$ is 5.8s and the constant of the governor speed-droop control $R$ is 1/17. For those generators with a rated power within 100 MW~200 MW, they are 8.1s and 1/20, respectively. For those generators with a rated power above 200 MW, they are 9.3s and 1/22, respectively. The amount of load damping $D$ is 2.5. The fraction of the power generator by the reheat turbine $F_H$ is 0.3. The average reheat time constant $T_R$ is 8. The power gain factor $K_m$ is 0.95. More detailed information about the test system can be found in [22].

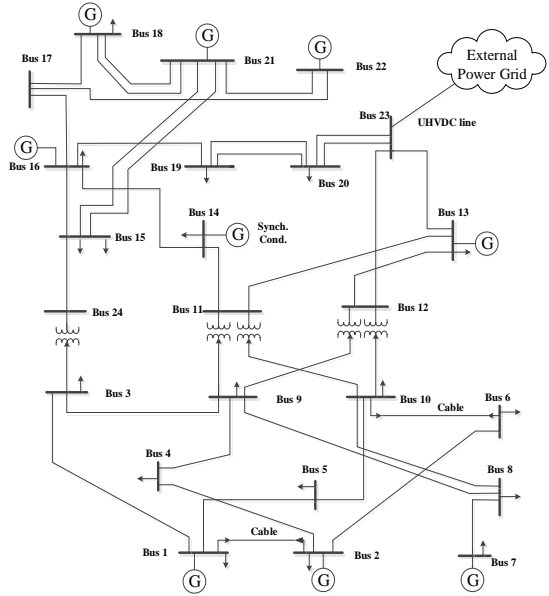

Fig. 12 A modified IEEE 24-bus system

## 6.1 Calculation of Threshold Power Loss

Based on the test system and the above parameters, it is calculated that $t_{nadir} = 3.72$s, and the threshold power loss $\Delta P_s$ is 351.90 MW, which makes the frequency drop to 49.5 Hz. The frequency drop curve is depicted in Fig. 13.

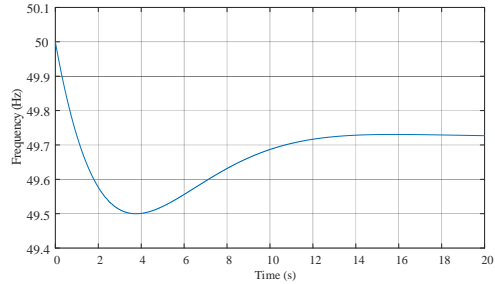

Fig. 13 The frequency drop in case of 351.9 MW power loss

## 6.2 Calculation of Load Shedding Conditions

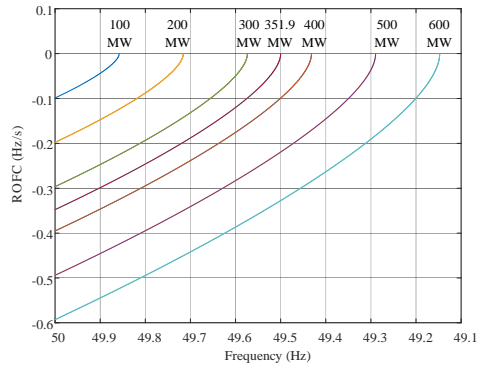

Fig. 14 Sensitivity analysis for the relationship between the ROCOF and frequency drop

A sensitivity analysis to determine the relationship between the ROCOF and frequency drop is conducted for different cases of power losses, i.e., 100 MW, 200 MW, 300 MW, 351.9 MW, 400 MW, 500 MW, and 600 MW, as shown in Fig. 14. It is seen that when at any frequency (e.g., 49.9 Hz), the ROCOF for a higher power loss is lower than that for a lower power loss, which validates that the ROCOF can be used an accurate index for detecting and responding to a contingency.

Thus, for each frequency measurement, if the frequency dropping rate is lower (i.e., dropping faster) than the ROCOF-frequency curve corresponding to the threshold power loss shown in Fig. 14, then the power loss could be larger than the threshold power loss, and load shedding is needed.

In practice, the ROCOF-frequency curve is discretized as shown in Table I.

Table 1. Specific load-shedding conditions of the smart outlets

| Frequency range | Load-shedding condition |
| --- | --- |
| [49.95, 50.00) | $ROFC < -0.3236$ Hz/s |
| [49.90, 49.95) | $ROFC < -0.2987$ Hz/s |
| [49.85, 49.90) | $ROFC < -0.2729$ Hz/s |
| [49.80, 49.85) | $ROFC < -0.2461$ Hz/s |
| [49.75, 49.80) | $ROFC < -0.2179$ Hz/s |
| [49.70, 49.75) | $ROFC < -0.1879$ Hz/s |
| [49.65, 49.70) | $ROFC < -0.1557$ Hz/s |
| [49.60, 49.65) | $ROFC < -0.1200$ Hz/s |
| [49.55, 49.60) | $ROFC < -0.0779$ Hz/s |
| [49.50, 49.55) | $ROFC < 0$ Hz/s |

Table 2. Specific switching-off conditions of the smart outlets

| Power loss | Random white noise in the measurements | Probability of deciding load shedding is needed |
| --- | --- | --- |
| 380 MW | 0 Hz | 1.000000 |
|  | 0.005 Hz | 0.999992 |
|  | 0.01 Hz | 0.978264 |
| 320 MW | 0 Hz | 0.000000 |
|  | 0.005 Hz | 0.001029 |
|  | 0.01 Hz | 0.117682 |

Simulations are conducted to check the accuracy of the load-shedding condition checking. Some random noise is intentionally added to each of the frequency measurements after a sudden power loss before checking if the frequency measurement meets the load shedding conditions in Table 1. The interval of the frequency measurements is 16 milliseconds. In order to improve the robustness of load-shedding conditions checking, the smart outlet will decide if load shedding is needed only when more than 15 frequency measurements have met the requirements in Table 1.

The simulation results are shown in Table 2. Two case studies are conducted for different power loss conditions. For each of the noises (i.e., 0 Hz, 0.005 Hz, and 0.1 Hz), simulations are conducted to generate 1 million frequency drop trajectories with random noise added R each point of the trajectory. At each point in the trajectory, we check to see whether load-shedding conditions are satisfied, based on which smart outlets are analyzed. For the case of 380 MW power loss which is larger than the threshold power loss, the smart outlet should decide that load shedding is needed. It is shown that the decisions of the smart outlets are very accurate as the probability (i.e., the probability of deciding load shedding is needed) is very high. For the case of 320 MW power loss, which is less than the threshold power loss, ideally, the smart outlet should decide that no shedding is needed. It is shown in Table 2 that the smart outlet decides that no load shedding is required in a majority of the simulations when there is no random noise or the noise is small (i.e., 0.005 Hz). In these two cases, the probability of wrong decisions is negligible. The probability of wrong decisions will only increase obviously only when the random white noise in the frequency is significantly high (i.e., 0.01 Hz).

*6.3 Local Estimation of Power Loss*

In this part, the local estimation of power loss using LSE and EKF is shown.

1) Performance of LSE

In order to show the performance of the LSE method for estimating the power loss, 1 million simulations are conducted assuming 500 MW (5 p.u.) power loss in the modified IEEE 24-bus system. For each simulation, a random noise within ±0.01 Hz is added to each of the frequency measurements, and the interval of the frequency measurements is 16 milliseconds. The simulation results are shown in Fig. 15.

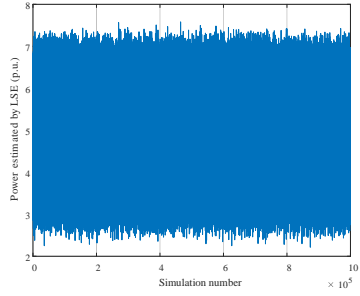

Fig. 15 Simulation results of the LSE method for power loss estimation

The results of power loss estimation are statistically analyzed in Table 3. It shows that the LSE method gives a reasonable but not very accurate result for the power loss right after the frequency disturbance event.

Table 3. Performance of the LSE for power loss estimation

| Range of power (p.u.) | Number of simulations | Ratio (%) |
| --- | --- | --- |
| [7, 8) | 1424 | 0.1424 |
| [6, 7) | 83166 | 8.3166 |
| [5, 6) | 369292 | 36.9292 |
| [4, 5) | 416360 | 41.6360 |
| [3, 4) | 125295 | 12.5295 |
| [2, 3) | 4463 | 0.4463 |

2) Performance of EKF

The power loss estimated by the LSE serves as the initial value for the EKF method. The EKF method tries to give a more accurate estimation of the power loss.

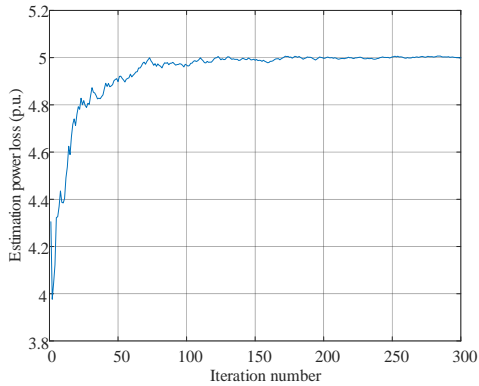

Fig. 16 Power loss evaluation curve of the EKF method

For each simulation in Fig. 15, the EKF method is adopted to further improve the accuracy of power loss estimation. A typical curve of the EKF results is shown in Fig. 16. The LSE estimates the power loss is 4.21 p.u., and this value serves as the initial value of the EKF method. With EKF, the power loss estimation gradually becomes very accurate. When the EKF converges, it will be very close to 5 p.u.

To demonstrate the statistical performance of the EKF method, for every simulation, the EKF analysis is performed for 40 iterations with an interval of 16 milliseconds. The results are shown in Fig. 17. By comparing Fig. 15 and Fig. 17, it is obvious that the EKF method further improves the accuracy of the power loss estimation results, as in general, the power estimation results come closer to 5 p.u.

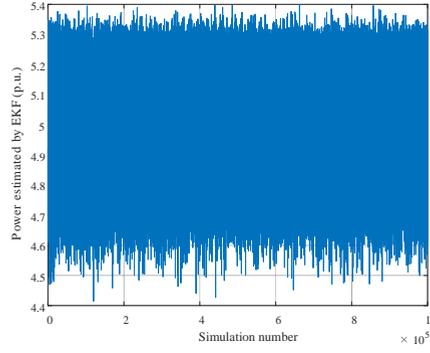

Fig. 17 Simulation results of the EKF method for power loss estimation

The results of power loss estimation based on the EKF method are statistically analyzed in Table 4. And it shows that 92.38% of the results have an error of less than 4%.

Table 4. Performance of the EKF for power loss estimation

| Range of power (p.u.) | Number of simulations | Ratio (%) |
| --- | --- | --- |
| [5.2, 5.4) | 36692 | 3.6692 |
| [5, 5.2) | 503811 | 50.3811 |
| [4.8, 5) | 419954 | 41.9954 |
| [4.6, 4.8) | 39086 | 3.9086 |
| [4.4, 4.6) | 457 | 0.0457 |

3) Robustness of EKF

Table 5. Performance of the EKF for power loss estimation with errors in system-level parameters

| Range of power (p.u.) | Number of simulations | Ratio (%) |
| --- | --- | --- |
| [5.6, 6) | 364 | 0.0364 |
| [5.4, 5.6) | 14662 | 1.4662 |
| [5.2, 5.4) | 135019 | 13.5019 |
| [5.0, 5.2) | 376247 | 37.6247 |
| [4.8, 5.0) | 357036 | 35.7036 |
| [4.6, 4.8) | 107807 | 10.7807 |
| [4.4, 4.6) | 8698 | 0.8698 |
| [4, 4.4) | 167 | 0.0167 |

In order to demonstrate the robustness of the EKF method when there are some errors in the system-level parameters, case studies are conducted with random noise in the system-level parameters including $G_1$, $G_2$, $\omega_n$, $\omega_r$, $\emptyset$. Specifically, a random noise within 5% is added to each of those parameters and 1 million simulations are conducted while a random noise within ±0.01 Hz still exists in each of the frequency measurements. The results of the EKF method after 40 iterations are shown in Table 5.

It can be seen from Table 5 that 72.33% of the power loss estimation results are within an error of less than 4%. This clearly demonstrates that the proposed EKF method can still accurately estimate the power loss even when there are random errors in the frequency measurements and in the system-level parameters.

*6.4 Experimental Validation of Power Loss Estimation*

In order to further validate the performance of the power loss estimation algorithm, hardware experiments are conducted. The dynamics of the system frequency after a sudden loss of 500 MW in the IEEE 24-bus system is simulated using Matlab. The simulated curve is loaded to the DSpace DS1202 MicroLabBox for real-time simulation. DSpace then sends the signals to a smart outlet for estimating the power loss. The system-level parameters related to the power loss estimation can be sent from the control center. Especially, 1% white noise is added to each of the points of the simulated curve.

The setup of the hardware experiment is shown in Fig. 18, and the results are obtained as shown in Fig. 19. The red line is the actual frequency curve of the IEEE 24-bus system. Before the sudden power loss, the power system operates in a normal state and the steady-state frequency is 50 Hz. Then, after a sudden power loss of 500 MW, frequency begins to drop rapidly.

The blue line shows the frequency measured by the smart outlet. As can be seen, the frequency measurement is accurate, but there can be some minor errors due to the 1% white noise in the input signal from the DSpace.

The abnormal frequency event detecting function is constantly working based on the frequency measurements. When the power system is working normally before the sudden power loss, it detects no abnormal event despite the fact that the noise in the input signal may cause some frequency measurement errors. However, 160 milliseconds after the sudden power loss, the event detection function detects that there is an abnormal event, which triggers the checking of the load-shedding conditions and the power loss estimation. For every new frequency measurement with an interval of 16 milliseconds, the load-shedding conditions in Table 1 are checked to see if they are satisfied; the load shedding conditions are seen as satisfied for 15 times after the event. In this case, the smart outlet will determine whether load shedding is needed.

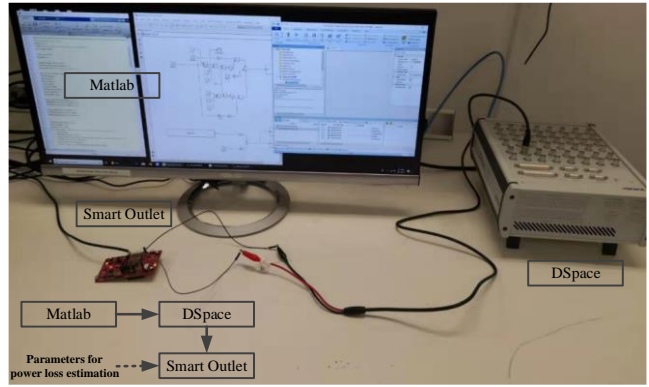

Fig. 18 Setup of the hardware experiment

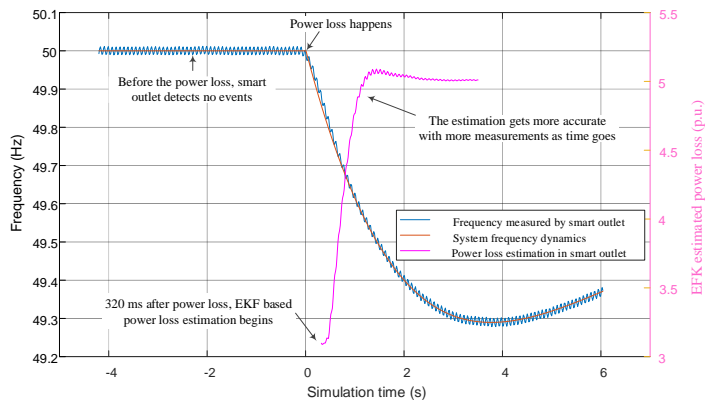

Fig. 19 Hardware experiment results

After the detection of an abnormal event, the LSE takes 10 frequency measurements with an interval of 16 milliseconds, estimates that the power loss is 3.10 p.u., and triggers the EKF-based power loss evaluation. The EKF takes the power loss estimated by LSE as its initial value, and for every new frequency measurement, it gives an updated result of power loss estimation, as shown as the pink curve in Fig. 19. It is shown that the EKF method can estimate the power loss fast and accurately, with increasing measurements obtained as time goes on.

In sum, this hardware experiment demonstrates that the proposed event detection method, load shedding condition checking method, and LSE and EKF based power loss methods are accurate and robust.

*6.5 Frequency Emergency Support for Power System*

Simulations are conducted to study the influence of the smart outlets' response on power system frequency emergency support. The load-shedding conditions can be usually satisfied when a sudden power loss occurs. As can be seen in Fig. 10, a smart outlet will switch off the appliance connected to it when its accumulated power sent from the cloud control center is less than the minimum load shedding requirement (i.e., the value of estimated power loss deducted by the threshold power loss).

Assuming there are 1 million smart outlets located in the IEEE 24-bus system, and the power of the appliance connected to a smart outlet is randomly located within [10 W, 1800 W]. The smart outlets are categorized into 1,000 load groups with 1,000 smart outlets in each group. Assume these load groups are ranked by their importance, shown as in Fig. 20.

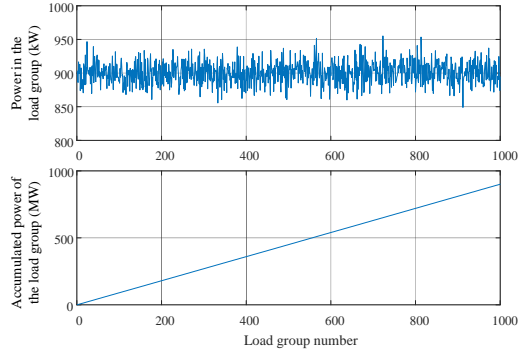

Fig. 20 Load groups and their accumulated power

1) Performance of Grid Sense for Frequency Emergency Support

The power loss estimation data of the 1 million smart outlets in Fig. 17 is applied to check the response of the smart outlets and their impact on the system frequency. The switching of each smart outlet is simulated based on the comparison of its accumulated power and estimated power loss deducted by the threshold power loss. In total, 164,793 smart outlets switched off, and the total load curtailment is 148.34 MW, which is very close to the load curtailment requirement 500-351.9=148.1 MW. This demonstrates that the proposed frequency control method is very accurate.

The actions of the smart outlets in each load group are shown in Fig. 21. It is noted the load groups are ranked from the least to the most important; thus, the smaller the load group number, the less important the loads connected to the smart outlets in the load group. As can be seen from Fig. 21, for the load groups 1~118, all the smart outlets in the load groups switch off. For the load groups 119~198, some smart outlets switch off and the others stay on. For the load groups larger than 199, no smart outlet switches off.

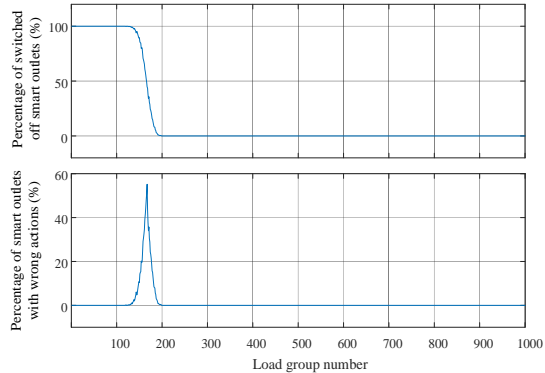

Fig. 21 Statistical analysis of the smart outlet actions in the load groups

The reason for the switching of the smart outlets is explained as follows. For the IEEE 24-bus system with a threshold power loss of 351.9 MW, if each smart outlet can accurately estimate the power loss (500 MW), ideally the smart outlets with an accumulated power below 148.1 MW should switch off, and the other smart outlets with an accumulated power larger than 148.1 MW should stay on. However, the power loss estimation by the smart outlets may not be 100% accurate as shown in Fig. 17. Thus, for the smart outlets with small values of accumulated power such as load groups 1~118, even if there might be some errors in the power loss estimation, the smart outlets will switch off as long as the estimated power loss deducted by the threshold power loss is larger than the value of accumulated power. Thus, the switching action of these smart outlets are always right, i.e., the least important smart outlets always switch off accurately.

Similarly, for the smart outlet with very large values of accumulated power such as in load groups 199~1000, the smart outlet will always stay on since the estimated power loss deducted by the threshold power loss is less than the value of accumulated power. Thus, the switching action of these smart outlets is also always right, i.e., the most important outlets always keep on accurately and their operation is not affected.

For the smart outlet with accumulated power values near 148.1 MW, they might not always take the right actions. It is checked that the accumulated power of load group 119 is 107.13 MW and the accumulated power of load group 198 is 178.12

MW. This is why the actions of the smart outlets in load groups 119~198 are not 100% correct. Further, it is analyzed that a load group with an accumulated power closer to 148.1 MW is more likely to have wrong actions. But fortunately, smart outlets below 148.1 MW might stay on while some outlets more than 148.1 MW might switch off. In such a case, these wrong actions compensate for each other, and the overall performance of all the smart outlets is still for the most part accurate.

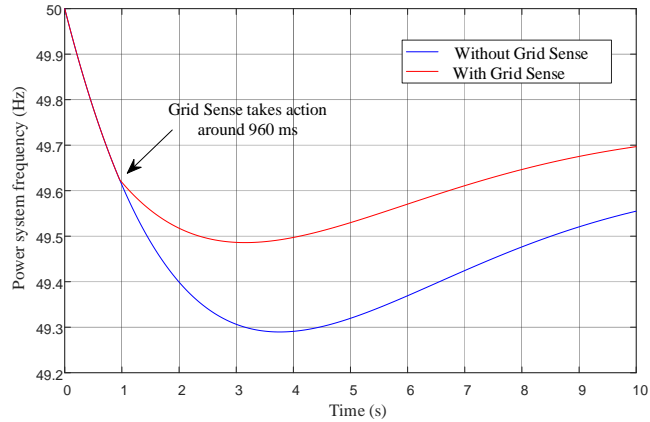

Fig. 22 Frequency regulation performance of Grid Sense

The switching time of the smart outlets depends on abnormal frequency disturbance event detection, load shedding condition checking, power loss estimation by LSE and EKF. Sometimes, the time taken by these functions may vary. Typically, the event detection is approximately 160 milliseconds, the LSE-based power loss estimation is 160 milliseconds, and EKF-based power loss estimation is 640 milliseconds. The load shedding conditions checking usually are less than 480 milliseconds, and it is conducted in parallel with the power loss estimation by LSE and EKF. Overall, the smart plugs typically switch off after around 960 milliseconds if a sudden power loss above the threshold occurs.

Considering the load shedding of the smart outlets, the power system frequency dynamics is shown in Fig. 22. It is shown here that the lowest frequency is around 49.5 Hz, which is the control objective of the Grid Sense system. Comparing the two cases without and with Grid Sense, it can be seen that the Grid Sense system can respond quickly and prevent the rapid dropping of the power system frequency.

2) Robustness of Frequency Emergency Support

It is demonstrated above that Grid Sense with a huge number of smart outlets can regulate the power system frequency accurately considering the noise in the frequency measurements. Additionally, it is also possible that the system-level parameters $G_1$, $G_2$, $\omega_n$, $\omega_r$, and $\emptyset$ used in the EKF-based power loss estimation might not be completely accurate. Thus, the 1,000,000 results of power loss estimation in Table 5 are adopted to simulate the performance of 1,000,000 smart outlets. It is found that 165,540 smart outlets switch off, and the total load shedding is 148.97 MW, which is still close to accurate.

This demonstrates that the Grid Sense system can work accurately even in case of random errors in the system-level parameters.

## 7. Conclusions and Future Work

This paper presents the cyber-physical architecture and the frequency control approaches of Grid Sense, which is a practical system designed to support the power system frequency using millions of distributed appliances under severe frequency disturbance contingencies. The Grid Sense system mainly deploys a local Δ*P*-estimation approach for frequency control, in which the control center estimates the parameters such as threshold power loss and send them to the smart outlets before the frequency disturbance event, while the smart outlets intelligently detect the frequency disturbance event, estimates the power loss, and makes switching decisions locally. Each part of the frequency control approach is illustrated in detail. Extensive simulations and hardware experiments are conducted to verify the accuracy and robustness of the proposed frequency control strategy.

In future work, more field tests will be conducted, and more types of randomness and errors will be considered and studied to further verify the accuracy, adaptivity, and robustness of Grid Sense.

**Acknowledgment**

This work was supported by the SGCC Science and Technology Program under project ''Distributed High-Speed Frequency Control under UHVDC Bipolar Blocking Fault Scenario'' under Grant SGGR0000DLJS1800934.